\documentclass{ws-ijmpd}
\usepackage[super,compress]{cite}
\usepackage{color}
\begin{document}

\markboth{Luis Rey D\'iaz-Barr\'on, Abraham Espinoza-García, S. P\'erez-Pay\'an, J. Socorro}
{Noncommutative Friedmann Equations in Effective LQC}

%
\catchline{}{}{}{}{}
%

\title{Noncommutative Friedmann Equations in Effective LQC}

\author{Luis Rey D\'iaz-Barr\'on\footnote{lrdiaz@ipn.mx}, Abraham Espinoza-Garc\'ia\footnote{aespinoza@ipn.mx} ~and S. P\'erez-Pay\'an\footnote{saperezp@ipn.mx}}

\address{Unidad Profesional Interdisciplinaria de Ingenier\'ia Campus Guanajuato del Instituto Polit\'ecnico Nacional.\\
	Av. Mineral de Valenciana \#200, Col. Fraccionamiento Industrial Puerto Interior, C.P. 36275, Silao de la Victoria, Guanajuato, M\'exico.\\}

\author{J. Socorro\footnote{socorro@fisica.ugto.mx}}
\address{Departamento de F\'{\i}sica, DCI-Campus Le\'on, Universidad de 
Guanajuato,\\
A.P. E-143, C.P. 37150,  Guanajuato, M\'exico.\\}

\maketitle

\begin{history}
\received{Day Month Year}
\revised{Day Month Year}
\end{history}

\begin{abstract}
In this work we construct a noncommutative version of the Friedmann equations in the framework of effective loop quantum cosmology, extending and applying the ideas presented in a previous proposal by some of the authors. The model under consideration is a flat FRW spacetime with a free scalar field. First, noncommutativity in the momentum sector is introduced. We establish the noncommutative equations of motion and obtain the corresponding exact solutions. Such solutions indicate that the bounce is preserved, in particular, the energy density is the same as in standard LQC. We also construct a noncommutative version of the modified Friedmann equations and argue that, as a consequence of noncommutativity, an effective potential arises. This, in turn, leads us to investigate the possibility of an inflationary era. Finally, we obtain the Friedmann and the Raychaudhuri equations when implementing noncommutativity in the configuration sector. In this case, no effective potential is induced.\end{abstract}

\keywords{Noncommutativity, Friedmann Equations, Effective Loop Quantum Cosmology}

\ccode{PACS numbers: 02.40.Gh, 98.80.Qc}


\section{Introduction}

It is believed that the correct description of the very early universe must come from a quantum theory of gravity since classical general relativity (GR) breaks down near the big-bang singularity. Loop quantum gravity (LQG) \cite{ashtekar-lqg, thiemann} is a leading candidate that aims to a full nonperturbative background independent quantization of GR.

Applying the methods and techniques of LQG to cosmological scenarios results in a framework known as loop quantum cosmology (LQC) \cite{bojowald-lrr, ashtekar-lqc}. That is, LQC results from implementing the quantization procedure developed in LQG to symmetry-reduced models which are cosmologically relevant. The endeavor to quantize cosmological models goes back to Wheeler and DeWitt with their pioneering work on canonical quantization \cite{DeWitt, Wheeler}. In the past two decades, LQC has had great success in quantizing minisuperspace models and has answered fundamental questions that the classical theory is not capable of doing. It has been shown that as a consequence of the underlying quantum geometry, the loop quantization of the FRW models exhibit a bounce, alleviating the problem of the cosmological singularity \cite{bojowald}. 

Effective equations based on a geometrical formulation of quantum mechanics can be obtained, which permit to study in a simple manner loop quantum corrections to cosmological models \cite{ashtekar-gfqm}. Numerical analyses using the effective dynamics for the FRW model with a free scalar field and for the anisotropic vacuum Bianchi I model have been studied, showing that this semiclassical effective scheme reproduces remarkably well the full quantum evolution \cite{Diener1, Diener2}.

On the other hand, the noncommutativity paradigm was revived at the end of the 20th century, mainly due to developments in string theory \cite{connes-douglas,seiberg-witten}, after a long period of lethargy from the first time it was introduced \cite{snyder}. Although the first works of noncommutative field theory were in connection with Yang-Mills theories, a short time later, noncommutative models of gravity were developed \cite{mofat}. Regrettably, all noncommutative theories of gravity share the common issue of being highly non-linear, which renders finding solutions to the noncommutative equations a very difficult task. It is believed that noncommutativity could play an important role in the evolution of the very early universe, which makes it a worthy candidate to study. Two seminal works led the pace concerning noncommutativity in the cosmological scenario. In the first one \cite{hugo}, the authors introduced the Moyal product of functions in the Wheeler-DeWitt equation for the Kantowski-Sachs (KS) cosmological model, and argued that noncommutative deformations affect the commutative fields, and that the effects of the full noncommutative theory of gravity should be reflected in the minisuperspace variables. The second one\cite{barbosa}, also studies noncommutative deformations for the KS model but already at the classical level, obtaining classical noncommutative equations of motion. Treading along these ideas, noncommutative Friedmann equations were obtained in Ref.~\refcite{wally-1}, when implementing a canonical deformation in the minisuperspace. It was also shown that the noncommutative contributions are only present to second order in the noncommutative parameter; and the phenomenological viability of the model was investigated. Other cosmological models following the spirit of these two influential works have also been investigated. \cite{Phantom_Wicho, Sinuhe_1, Sinuhe_2, Aguero}.

Considering the simplicity achieved by the effective scheme of LQC in describing the full quantum evolution of the FRW model (see   Ref.~\refcite{Diener1} and Ref.~\refcite{Diener2}), it is rather  natural to employ such a framework as a first approximation to probe the coupling of noncommutativity to LQC. Furthermore, as stated above, noncommutativity has already been implemented in a simple way in the classical phase space of different cosmological models (see, for instance, Ref.~\refcite{barbosa} and Ref.~\refcite{wally-1}). Such effective noncommutative LQC could be interpreted as a symmetry-reduced case of the more general framework constructed in  Ref.~\refcite{kober}. This line of thought led to some of the authors of the present investigation to propose a simple noncommutative framework of the effective LQC for the flat FRW model with a free scalar field\cite{primogenito}. The main conclusion was that a canonical noncommutativity in the momentum sector is more compatible with LQC than its configuration sector counterpart, in the sense that key features of LQC are better retained when noncommutativity is introduced among the momentum variables (e.g., the bounce and the energy density profile). 

In the present work, we extend and apply the construction made in Ref.~\refcite{primogenito}, and derive a noncommutative version of the Friedmann equations for each sector of phase space. Analytical solutions of the noncommutative equations of motion (EOM) in the momentum sector are obtained (this could not be achieved in Ref.~\refcite{primogenito}). It is found that the volume of the flat FRW universe preserves the bounce, and that after the bounce it has a slower growth than its commutative counterpart. Also, the energy density is explicitly calculated and shown not to depend on the noncommutative parameter. These conclusions were also reached in Ref.~\refcite{primogenito}, but different arguments were given. Finally, we argue that an effective (scalar field) potential is induced due to the presence of noncommutativity, giving rise to an accelerated universe; a compelling reason to investigate a possible inflationary era.
 
The manuscript is organized as follows: In section II we briefly introduce the effective scheme of loop quantum cosmology for the flat FRW model with a free scalar field. Section III is devoted to implementing a canonical noncommutativity at the effective scheme of LQC, and to studying some of its consequences via solutions to the EOM. Finally, section IV is dedicated to discussion and outlook.
\section{Effective Loop Quantum Cosmology}
LQG is a canonical quantization of gravity whose fundamental components are the $SU(2)$ Ashtekar-Barbero connection $A_a^i$ and the densitized triad $E^a_i$ (labels $a$ and $i$ denote space and internal indices, respectively), which are two conjugate variables encoding the curvature and spatial geometry, respectively. Similarly, LQC is a canonical quantization, based on the techniques used in LQG, of cosmologically relevant symmetry-reduced models. In LQC, the imposition of homogeneity and isotropy allows the connection and triad to be described by parameters $c$ and $p$, respectively; i.e., the phase space structure is simplified. The canonically conjugate pair $(c,p)$ satisfy $\{c,p\}=8\pi G\gamma/3$, where $G$ is Newton's gravitational constant and $\gamma$ is the dimensionless Barbero-Immirzi parameter \cite{ashtekar-lqc}. For the flat FRW model, this new set of variables is related to the metric variables as $c=\gamma\dot{a}$ and $p=a^2$, where $a$ is the scale factor of the universe. 

If one introduces the variables $\beta$ and $v$, defined in terms of $c$ and $p$ by
\begin{equation}
\beta=\frac{c}{\sqrt p}, \quad v=p^{3/2},
\end{equation}
the classical Hamiltonian can be written as
\begin{equation}
\mathcal{H}_{cl}=N\left[\frac{-3v}{8\pi G\gamma^2}\beta^2+\frac{p_{\phi}^2}{2v}\right],
\end{equation}
 where $N$ is the lapse function and $\{\beta,v\}=4\pi G\gamma$.

In the Hamiltonian formulation for homogeneous and isotropic models, the only relevant constraint is the Hamiltonian constraint, whose vanishing gives the physical solutions. Upon quantization, the Hamiltonian constraint operator is obtained by promoting the holonomies and the triads to the corresponding operators. As a result, the description of LQC is given by a discrete difference equation, signaling that the underlying geometry of LQC is discrete \cite{bojowald, Ashtekar-Bojo-Lewa}. 

However, an effective framework on classical phase space can be constructed using semiclassical states, which had been shown to be in excellent agreement with the quantum dynamics \cite{ashtekar-gfqm, Ashtekar-numerical, ashtekar-frw}. In this effective frame the equations that describe the spacetime are a modified version of the classical Friedmann equations. The effective Hamiltonian for the flat FRW with a free scalar field as the matter content, is given by \cite{robustness}
\begin{equation}
\mathcal{H}_{\mathrm{eff}}=N \left[-\frac{3v}{8\pi G\gamma^{2}\lambda^{2}}\sin^{2}(\lambda\beta)+\frac{p^{2}_{\phi}}{2v}\right],
\label{ham2}
\end{equation}
where $\lambda^{2}=4\sqrt{3}\pi\gamma\ell^{2}_{p}$ is the smallest eigenvalue of the area operator in the full LQG and  $\ell_p$ is the Planck length \cite{thiemann}. From the effective Hamiltonian (\ref{ham2}), we can calculate the equation for $v$, through Hamilton's equation:
\begin{equation}
\dot v=-4\pi G\gamma\frac{\partial\mathcal{H}_{\mathrm{eff}}}{\partial\beta}=
\frac{3v}{\gamma\lambda}\sin(\lambda\beta)\cos(\lambda\beta)\label{V_conm},
\end{equation}
where we have set the lapse function equal to one, a choice that we will keep in the remainder of the manuscript (a careful analysis pertaining to this particular gauge is presented in Ref.~\refcite{robustness}). Moreover, the vanishing of the Hamiltonian constraint (which the physical solutions must satisfy) $\mathcal{H}_{eff}\approx 0$, implies
\begin{equation}
\frac{3}{8\pi G\gamma^2\lambda^2}\sin^2(\lambda\beta)=\frac{p_{\phi}^2}{2v^2}.\label{constriction}
\end{equation}
Combining equations (\ref{V_conm}) and (\ref{constriction}), it is straight forward to construct the modified Friedmann equation, $\mathrm{H}=\dot v/3v$, \cite{Singh2006,Taveras, Vandersloot}
\begin{equation}
\mathrm{H}^{2}=\frac{8\pi G}{3}\rho\left(1-\frac{\rho}{\rho_{max}}\right),
\label{eff-FRW}
\end{equation}
where $\rho=(\dot\phi)^2/2=p^{2}_{\phi}/2v^2=(3/8\pi G\gamma^{2}\lambda^{2})\sin^{2}(\lambda\beta)$ and $\rho_{max}$ is the maximum value that $\rho$ can take, that is, $\rho_{max}=3/8\pi G\gamma^{2}\lambda^{2}$. It can be seen from Eq.~\ref{eff-FRW} that the Friedmann equation incorporates holonomy corrections due to LQG, manifested in the $\rho^2$ term, which enables an avenue to investigate the role of nonperturbative quantum corrections. Also, the turning points of the volume function occur at $\beta=\pm{\pi}/({2\lambda})$, which correspond to a bounce. In the limit $\lambda\rightarrow0$ (no area gap), the ordinary Friedmann equation, $\mathrm H^2=(8\pi G/3)\rho$, is recovered.

Similarly, the Raychaudhuri equation can be obtained from the field equation for $\beta$, which yields \cite{Taveras}
\begin{eqnarray}
\frac{\ddot{a}}{a}&=&-\frac{16\pi G}{3}\rho\left(1-\frac{5}{2}\frac{\rho}{\rho_{max}}\right).
\label{eff-RE}
\end{eqnarray}
\section{Noncommutative Friedmann Equations}
In standard canonical quantum cosmology, the Wheeler-DeWitt (WDW) equation is responsible for the description of quantum evolution. In this framework, the gravitational and matter fields have been reduced to a finite number of degrees of freedom due to the underlying symmetry of the model under consideration. This results in a finite dimensional phase space, the configuration sector of which is termed {\it minisuperspace} \cite{ryan}. An alternative approach to study quantum mechanical effects is to introduce deformations to the (algebra of functions of) phase space (this particular way of quantizing is part of a complete and consistent type of quantization known as deformation quantization \cite{flato}). The first ideas in connection of deformed minisuperspace were done in noncommutative cosmology \cite{hugo}, in order to incorporate an effective noncommutativity. Therefore studying cosmological models in deformed phase space could be interpreted as studying quantum effects to cosmological solutions \cite{Vakili,Malekolkalami}. According to the phase space deformation procedure, the deformation is coded in the Moyal brackets $\{f,g\}_{\alpha}=f\star_{\alpha}g-g\star_{\alpha}f$, where the usual product of functions is replaced by the Moyal product
$(f\star g)(x)=\exp\left[\frac{1}{2}\alpha^{ab}\partial_a^{(1)}\partial_b^{(2)}\right]f(x_1)g(x_2)\vert_{x_1=x_2=x}$,
such that
\begin{eqnarray}
\alpha =
\left( {\begin{array}{cc}
 \theta_{ij} & \delta_{ij}+\sigma_{ij}  \\
- \delta_{ij}-\sigma_{ij} & \eta_{ij}  \\
 \end{array} } \right),
\end{eqnarray}
where $\theta_{ij}$ and $\eta_{ij}$ are antisymmetric and represent the noncommutativity between the coordinates and the momenta, respectively; and $\sigma_{ij}=-1/8(\theta_{i}^k\beta_{kj}+\beta_{i}^k\theta_{kj})$. The resulting $\alpha$-deformed algebra for the phase space is
\begin{equation}
\{x_i,x_j\}_{\alpha}=\theta_{ij}, \;\{x_i,p_j\}_{\alpha}=\delta_{ij}+\sigma_{ij},\; \{p_i,p_j\}_{\alpha}=\eta_{ij},\label{alg}
\end{equation}
It follows that there is an alternative to derive a similar algebra to Eq.~(\ref{alg}). Consider the following transformation on the phase space variables $\{x,y,p_x,p_y\}$,
\begin{eqnarray}\nonumber
\hat{x}=x+\frac{\theta}{2}p_{y}, \qquad \hat{y}=y-\frac{\theta}{2}p_{x},\\
\hat{p}_{x}=p_{x}-\frac{\eta}{2}y, \qquad \hat{p}_{y}=p_{y}+\frac{\eta}{2}x, \label{nctrans1}
\end{eqnarray}
with $\{x,y,p_x,p_y\}$ satisfying the usual Poisson algebra. The new variables fulfill a deformed algebra   
\begin{equation}
\{\hat{y},\hat{x}\}=\theta,\; \{\hat{x},\hat{p}_{x}\}=\{\hat{y},\hat{p}_{y}\}=1+\sigma,\; \{\hat{p}_y,\hat{p}_x\}=\eta,\label{dpa}
\end{equation}
where $\sigma=\theta\eta/4$, and we have used $\theta_{ij}=-\theta\epsilon_{ij}$, $\beta_{ij}=\beta\epsilon_{ij}$ ($\epsilon_{ij}$ being the Levi-Civita symbol). We can see that relations above are of the same form as Eq.~(\ref{alg}), however, the brackets in Eq.~(\ref{dpa}) are usual Poisson brackets, whereas the brackets in Eq.~(\ref{alg}) are $\alpha$-deformed Poisson brackets. 

In practice, it is easier to work with usual Poisson brackets (\ref{dpa}) than with $\alpha$-deformed Poisson brackets (\ref{alg}). Let $\mathcal{H}(x,y,p_x,p_y)$ be the Hamiltonian function characterizing a classical mechanical system in a Darboux chart. The equations of motion for $\{x,y,p_x,p_y\}$ (obtained as in a Darboux chart) employing as Hamiltonian the function $\mathcal{H}^{nc}(x,y,p_x,p_y)=\mathcal{H}(\hat{x}(x,y,p_xp_y),\hat{y}(x,y,p_xp_y),\hat{p}_x(x,y,p_xp_y),\hat{p}_y(x,y,p_xp_y))$ involve noncommutative terms which are related to the algebra (\ref{dpa}). Solutions to these equations are appropriately termed noncommutative solutions, they reduce to the commutative ones in the limit where the noncommutative parameters go to zero \cite{barbosa, wally-2}. We will follow this ``shifted variables'' prescription.

\subsection{NC in the momentum sector of effective lqc}\label{section_mom}
Considering the above discussion, we have two choices to formulate the deformed theory. The first is to apply the $\alpha$-deformed algebra  (\ref{alg}) (via $\alpha$-deformed Poisson brackets) to a system characterized by a canonical Hamiltonian $\mathcal{H}$, resulting in a deformed (noncanonical) structure of the EOM. 

The second one is to introduce ``shifted variables'' defined by transformations (\ref{nctrans1}), where the structure of the EOM is the usual canonical one (a deformation of Poisson brackets is not performed). Here the Hamiltonian $\mathcal{H}^{nc}$ has the same functional form as $\mathcal{H}$ but is valued in the variables which fulfill the algebra (\ref{dpa}). As already stated, in this work we are going to stick with choice number two (this approach has been used in different cosmological scenarios\cite{Phantom_Wicho,Sinuhe_1,Sinuhe_2,Vakili,Malekolkalami}).

We start by implementing a deformed algebra, analogous to Eq.~(\ref{dpa}), but in the phase space spanned by the variables $\{\beta, \phi, v, p_\phi\}$ of the effective LQC scheme. For that purpose, let us consider the deformed algebra in the momentum sector
\begin{equation}
\{\beta^{nc},v^{nc}\}=4\pi G\gamma,\quad \{v^{nc},p^{nc}_{\phi}\}=\eta, \quad\{\phi^{nc},p^{nc}_{\phi}\}=1,
\label{ncm-free}
\end{equation}
with the remaining brackets being zero. The above relations can be implemented working with the shifted variables
\begin{equation}
\beta^{nc}=\beta, \quad \phi^{nc}=\phi, \quad v^{nc}=v+a\eta\phi, \quad p_{\phi}^{nc}=p_\phi+b\eta\beta,
\label{nc-rel2}
\end{equation}
where $a$ and $b$ satisfy the relation $a-4\pi G\gamma b=1$.
As stated above, we are going to construct the deformed theory starting with the effective Hamiltonian, which is formally analogous to (\ref{ham2}), but constructed with variables that obey the algebra given by Eq.~(\ref{ncm-free}), resulting in
\begin{equation}
\mathcal{H}^{nc}_{\mathrm{eff}}=-\frac{3v^{nc}}{8\pi G\gamma^{2}\lambda^{2}}\sin^{2}(\lambda\beta)+\frac{\left(p_{\phi}^{nc}\right)^{2}}{2 v^{nc}}.
\label{nc2-frw-ham}
\end{equation}
The noncommutative effective field equations are
\begin{eqnarray}\label{edm_momentos}
\dot{\beta}&=&4\pi G\gamma\frac{\partial \mathcal{H}^{nc}_{\mathrm{eff}}}{\partial v}=-\frac{3}{2\gamma\lambda^{2}}\sin^{2}(\lambda\beta),\nonumber\label{edm_mom_beta}\\ 
\dot{v}&=&-4\pi G\gamma\frac{\partial \mathcal{H}^{nc}_{\mathrm{eff}}}{\partial\beta}=\frac{3 v^{nc}}{\gamma\lambda}\sin(\lambda\beta)\cos(\lambda\beta)-\frac{4\pi G\gamma b\eta p_{\phi}^{nc}}{v^{nc}},\nonumber\label{edm_mom_vol}\\ 
\dot{\phi}&=&\frac{\partial \mathcal{H}^{nc}_{\mathrm{eff}}}{\partial p_{\phi}}=\frac{p_{\phi}^{nc}}{v^{nc}},\label{edm_mom_fi}\nonumber\\ 
\dot{p}_{\phi}&=&-\frac{\partial \mathcal{H}^{nc}_{\mathrm{eff}}}{\partial\phi}=\frac{3a\eta}{8\pi G\gamma^{2}\lambda^{2}}\sin^{2}(\lambda\beta)+a\eta\frac{\left(p_{\phi}^{nc}\right)^{2}}{2\left(v^{nc}\right)^{2}}\label{{edm_mom_pfi}},
\end{eqnarray}
and in the limit $\eta\rightarrow0$, we recover the commutative field equations.

\begin{figure}
  \includegraphics[width=\linewidth]{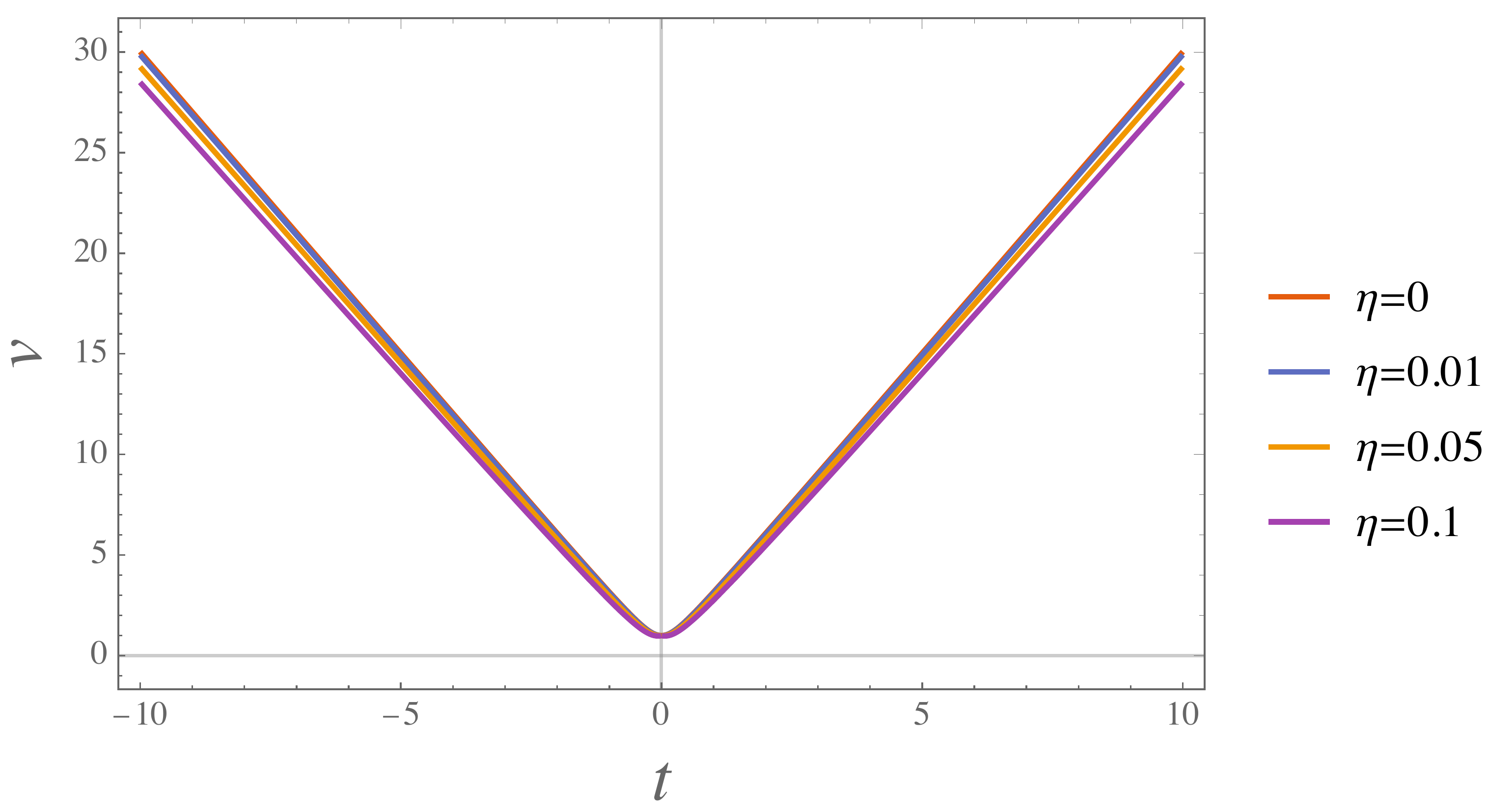}
  \caption{\label{figura1a}{\small After the bounce the volume grows slower as $\eta$ takes larger values}}
\end{figure}

\begin{figure}
  \includegraphics[scale=0.45]{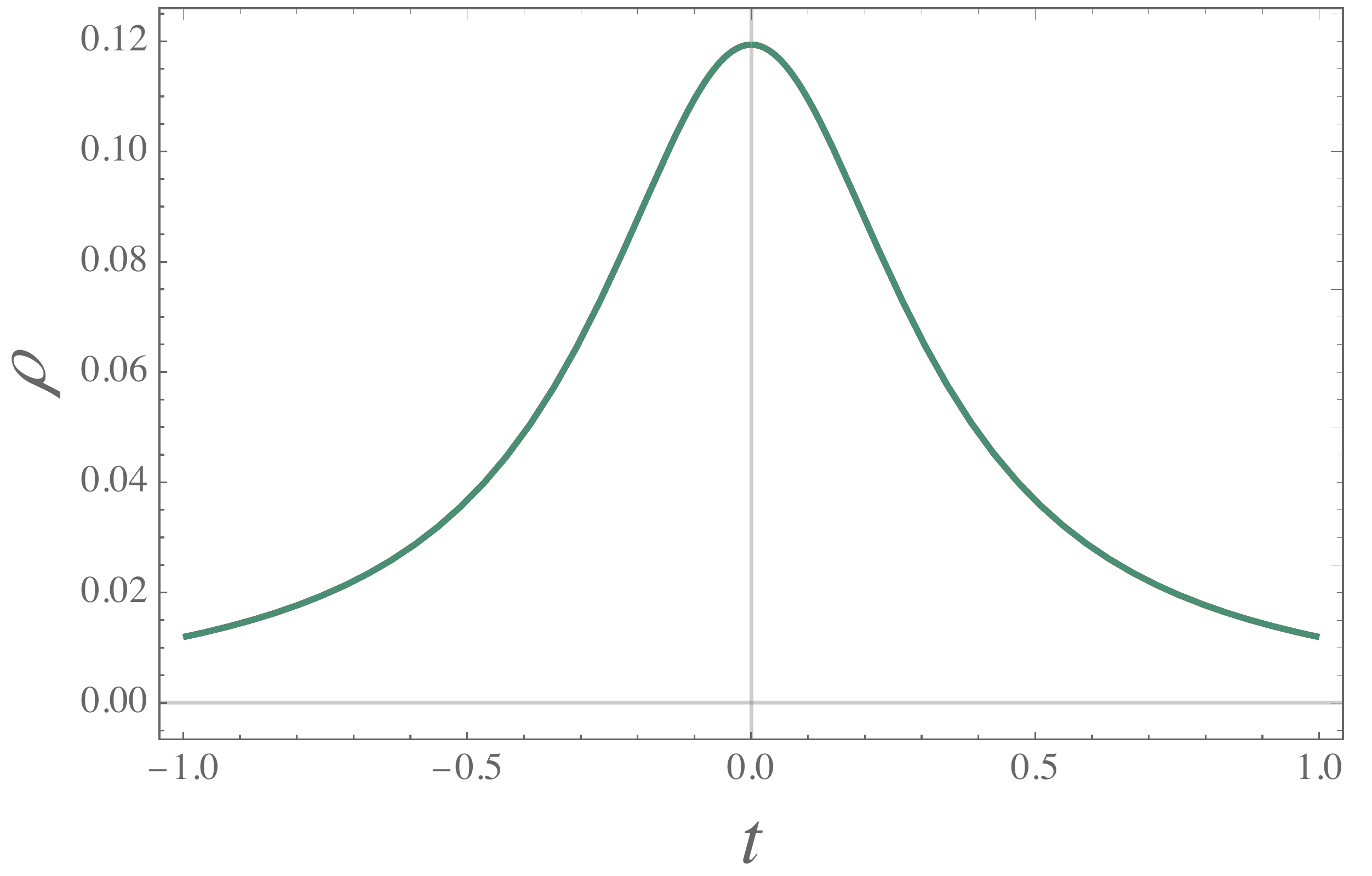}
  \caption{\label{figura1b}The energy density profile is the same as in standard LQC}
\end{figure}

\begin{figure}
  \includegraphics[width=\linewidth]{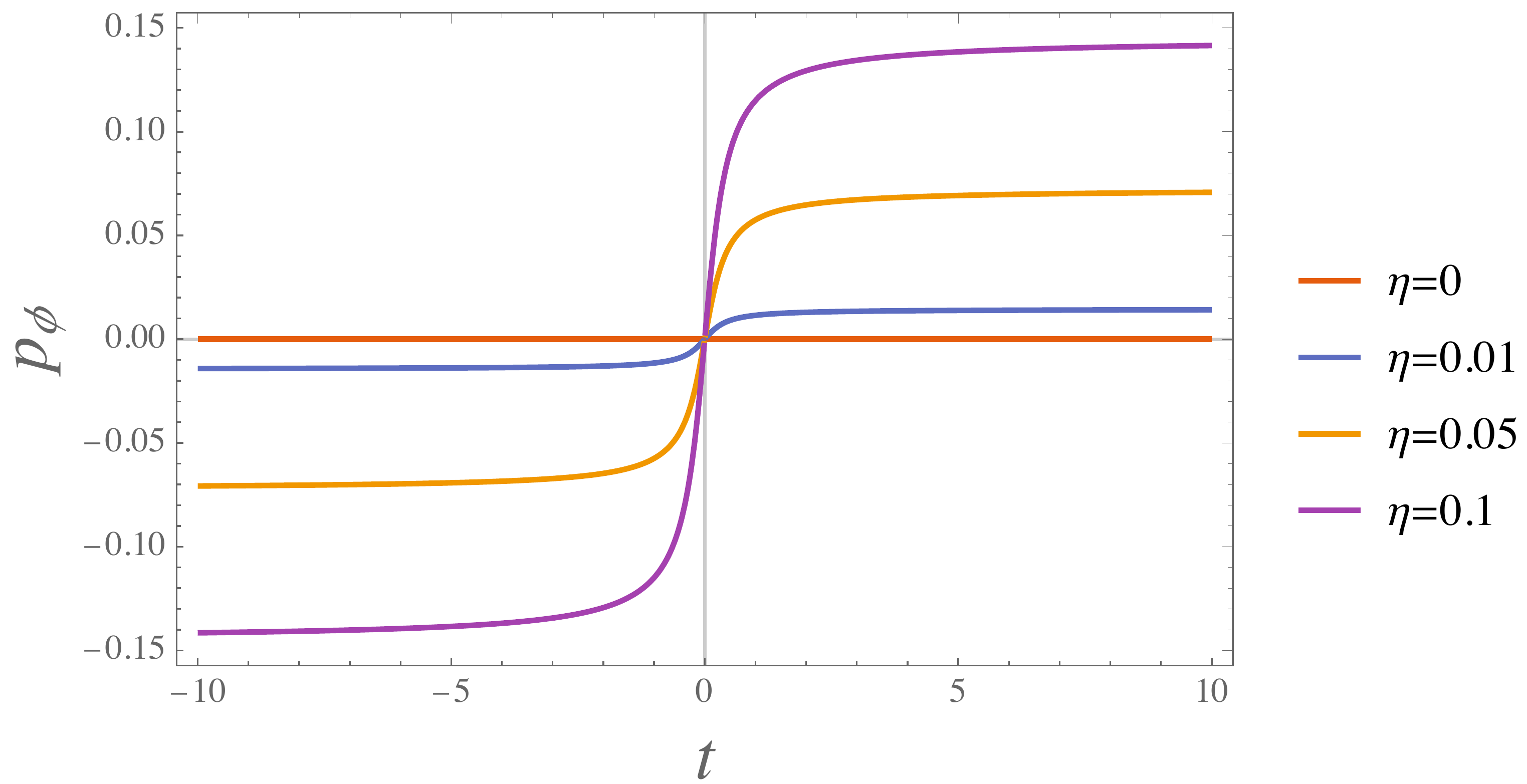}
  \caption{\label{figurapphi} Behavior of  $p_\phi$ for different values of the noncommutative parameter (where we have set $C_2=0$).}
\end{figure}

The equation for $\beta(t)$ is the same as in standard LQC, which enables us to obtain analytical solutions of the remaining equations of motion. The volume for flat FRW model in this setup now is given by
\begin{equation}
v(t)=C_1\sqrt{9t^2+\gamma^2\lambda^2}-\frac{\eta(4\pi\gamma bG-a)}{\lambda\sqrt{12\pi G}}\tan^{-1}\left(\frac{3t}{\gamma\lambda}\right)+\frac{a\gamma\eta}{\sqrt{12\pi G}}\frac{\log(\sqrt{9t^2+\gamma^2\lambda^2}+3t)}{\sqrt{9t^2+\gamma^2\lambda^2}},
\end{equation}
$C_1$ is a constant coming from integration. We can observe that the volume gets modified  by the presence of noncommutativity, in contrast to the solution for $\beta$. This modification causes the volume (after the bounce) of the FRW universe to have a slower growth, as the noncommutative parameter $\eta$ takes larger values, as shown in Fig.~\ref{figura1a}.
For $\phi(t)$, we have
\begin{equation}
\phi(t)=\frac{\gamma}{\sqrt{12\pi G}}\log{(\sqrt{9t^2+\gamma^2\lambda^2}+3t)},
\end{equation}
where we can see that $\phi(t)$ does not depend on the noncommutative parameter. With these solutions at hand, the energy density profile $\rho=(\dot\phi)^2/2$ can be obtained. Fig.~\ref{figura1b}, shows that the behavior for $\rho$ is the same as in standard LQC as well as the maximum value that can take. The same conclusion was made in Ref.~\refcite{primogenito}, but different arguments were given.
Finally, the solution for $p_\phi(t)$ is given by
\begin{equation}
\label{solpp}
p_\phi(t)=C_2+\frac{a\eta}{4\pi G\gamma\lambda}\arctan\left(\frac{3t}{\gamma\lambda}\right),
\end{equation}
where $C_2$ is an integration constant. In the limit $\eta\to0$, the commutative solution is obtained. From Eq.~(\ref{solpp}) we can see that $p_\phi$ incorporates noncommutative corrections, and went from a constant value to a one that increases as $\eta$ increases, as depicted in Fig.~\ref{figurapphi}.
From the field equation for the volume and the noncommutative effective Hamiltonian constraint ($\mathcal{H}_{eff}^{nc}\approx 0$), we are able to construct a Friedmann equation that incorporates the quantum corrections, as well as the noncommutative corrections, which reads
\begin{equation}\label{Friedmann_NC_Mom}
\mathrm{H}^2=\frac{8\pi G}{3}\rho\left(1-\frac{\rho}{\rho_{max}}\right)\left(\frac{1-\zeta_{max}}{1-\xi_{max}}\right)^2,
\end{equation}
where
\begin{equation}
\zeta_{max}=\frac{\eta}{\sqrt{12\pi G}C_3\gamma\lambda}
\left[(1+a)\sqrt{\frac{\rho}{\rho_{max}-\rho}}+\arctan\left(\sqrt{\frac{\rho_{max}-\rho}{\rho}}\right)\right],\label{zeta_max}
\end{equation}
and
\begin{multline}
\xi_{max}=\frac{\eta}{\sqrt{12\pi G}C_3\gamma\lambda}\left\{\arctan\left(\sqrt{\frac{\rho_{max}-\rho}{\rho}}\right)\right.\\
+ \left. a\sqrt{\frac{\rho}{\rho_{max}}}\ln\left[\frac{\gamma\lambda}{\sqrt\rho}\left(\sqrt\rho+\sqrt{\rho_{max}-\rho}\right)\right]\right\}\label{chi_max},
\end{multline}
where $C_3$ is an integration constant. 

Despite having at our disposal the solutions to the noncommutative EOM, we were not able to write the Raychaudhuri equation in a compact form. However, it is easy to check that the equation
\begin{equation}
\frac{\ddot a}{a}=\dot{\mathrm H}+\mathrm{H}^2,\label{aceleracion_momentos}
\end{equation}  
reduces to Eq.~(\ref{eff-RE}) for $\eta\to0$.

\begin{figure}
  \includegraphics[width=\linewidth]{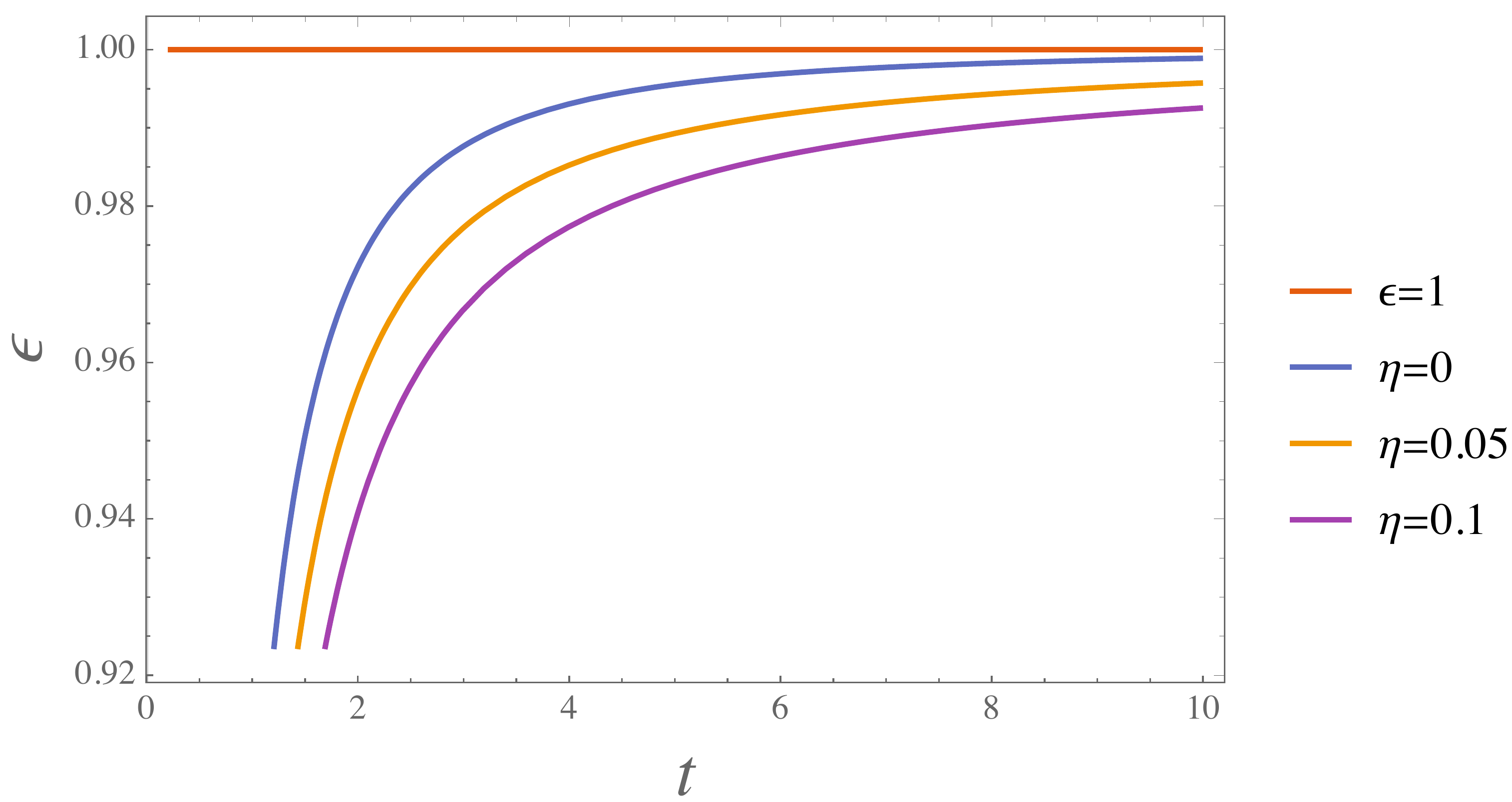}
  \caption{\label{epsilon}Dynamical behavior of the slow-roll parameter for different values of the noncommutative parameter $\eta$.}
\end{figure}

\begin{figure}
  \includegraphics[width=\linewidth]{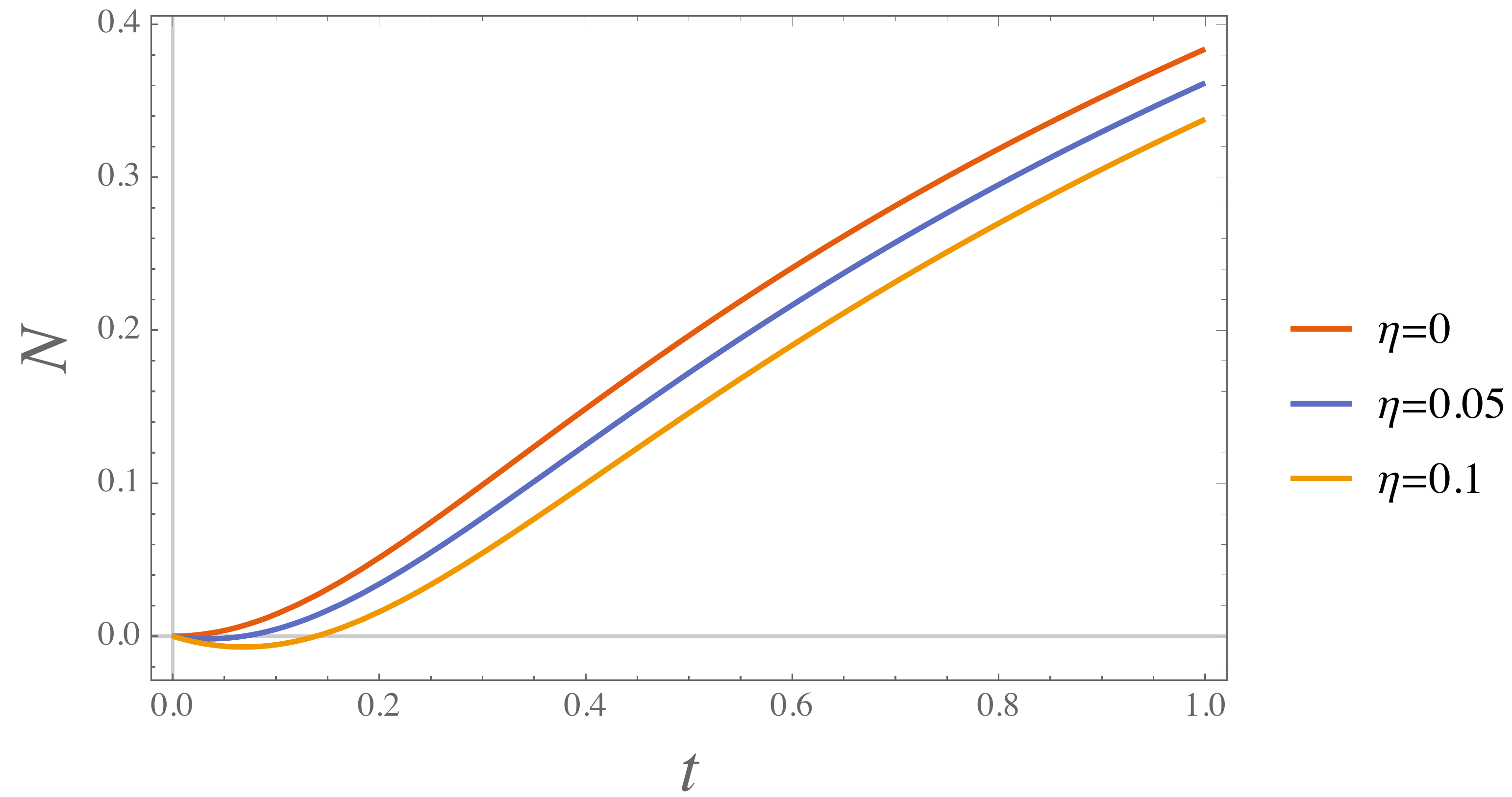}
  \caption{\label{num_efoldings} Number of e-foldings as the noncommutative parameter takes on different values.}
\end{figure}

We can interpret the dependency of the noncommutative Hamiltonian (\ref{nc2-frw-ham}) on the scalar field $\phi$ as the emergence of an effective potential due to noncommutativity. Establishing that identification, and since solutions for the field equations are now available, we investigate if this effective scalar field is capable to originate an inflationary epoch. We start by rewriting Eq.~(\ref{aceleracion_momentos}) in the following form  
\begin{equation}
\label{Hub_epsilon}
\frac{\ddot a}{a}=\mathrm{H}^2\left(1-\epsilon\right),
\end{equation}
where $\epsilon\equiv-\dot{\mathrm H}/\mathrm{H}^2$ is the Hubble slow-roll parameter. It is well known that in an inflationary era the slow-roll parameter satisfies: $\epsilon<1$. Fig.~\ref{epsilon} shows the dynamical behavior of $\epsilon$, and it can be seen that the slow roll parameter meets this condition, nonetheless, the fulfillment of this requirement is not sufficient to guarantee inflation. Also, we can observe that for early times, the universe has an accelerated behavior, that is, $\ddot a>0$. In order to have a deeper understanding of this analysis, we calculate the number of e-foldings, $N(t)$, which are given by
\begin{equation}
N(t)=\int_{t}^{t_{max}}H(t^{\prime})dt^{\prime}.
\end{equation}
Fig.~\ref{num_efoldings} shows the number of e-foldings for the model, and it can be seen that for early times, the e-foldings are less than one. In the literature, it is well documented that a universe that undergoes an inflationary phase must have at least $N(t)\approx 60$ \cite{mukhanov}, thus, our model is far from an inflationary stage.

\subsection{NC in the configuration sector of effective lqc}
For completeness, we derive the noncommutative version of the Friedmann equations in the configuration sector, something that could not be achieved in Ref.~\refcite{primogenito}. We start by recalling the deformed Hamiltonian in the configuration sector, which reads
\begin{equation}\label{nc-frw-ham}
\mathcal{H}^{nc}_{\mathrm{eff}}=-\frac{3v}{8\pi G\gamma^{2}\lambda^{2}}\sin^{2}(\lambda\beta^{nc})+\frac{p^{2}_{\phi}}{2v},
\end{equation}
from which the equations of motion are calculated using Hamilton's equations and whose solutions are given by \cite{primogenito}
\begin{eqnarray}
\label{edm_config}
\beta(t)&=&\frac{1}{\lambda}\mathrm{arccot}\left(\frac{3t}{\gamma\lambda}\right)-a\theta p_{\phi},\nonumber\label{sol_nc_beta}\\
v(t)&=&C_1^{\prime}\sqrt{\gamma^{2}\lambda^{2}+9t^{2}},\nonumber\\
\phi(t)&=&C_2^{\prime}+\frac{p_{\phi}}{3C_1}\mathrm{log}\left(3t+\sqrt{\gamma^2\lambda^2+9t^2}\right)- \frac{aC^2_1\theta\sqrt{\gamma^2\lambda^2+9t^2}}{4C_1\pi G\gamma},\nonumber\\
p_{\phi}&=&C_3^{\prime},\label{sol_nc_pfi}
\end{eqnarray}
where $C_1^{\prime}$, $C_2^{\prime}$ and $C_3^{\prime}$ are integration constants. Considering the last set of equations (\ref{edm_config}), we can construct the energy density profile, which is now given by
\begin{equation}
\label{rhoncm}
\rho(t)=\rho_{max}\left(1-\frac{3ap_\phi\theta t}{\gamma}\right)^2\left(1+\frac{9t^2}{\gamma^2\lambda^2}\right)^{-1},
\end{equation}
it can be seen that this quantity gets modified by the presence of noncommutativity. In the limit $\theta\to 0$ we recover the standard LQC energy density profile. The maximum value of Eq.~(\ref{rhoncm}) is $\rho_{max}(1+a^2\theta^2\lambda^2p_{\phi}^2)\equiv\rho_{max}^{nc}$ and it is reached at $t=-\frac{1}{3}a\theta\gamma\lambda^2p_\phi$. It is important to emphasize that the bounce continues to happen at $t=0$, as the equation for $v(t)$ indicate. The behavior of $\rho_{max}^{nc}$ is the same as the one reported already in Ref.~\refcite{primogenito}. The shape of the density profile differs more and more from the one in standard LQC as $\theta$ takes on larger values. From Eqns. (\ref{edm_config}) and (\ref{rhoncm}) we find that the modified Friedmann equation with noncommutative corrections takes the form
\begin{equation}
\mathrm{H}^2=\frac{8\pi G}{3}\rho\left(1-\frac{\rho}{\rho_{max}}\right)\left(\frac{1+\zeta_\theta}{\xi_\theta{}^2}\right),\label{Friedmann_NC_Config}
\end{equation}
and the modified acceleration equation incorporating noncommutative corrections is
\begin{equation}
\frac{\ddot a}{a}=-\frac{16\pi G}{3}\rho\left\lbrace1-\frac{5}{2}\left(\frac{\rho_{max}^{nc}-\rho_{max}}{\rho}-1\right)^2\frac{1}{\chi_\theta}\right\rbrace
 \left(\frac{\rho_{max}^{nc}-\rho_{max}}{\rho}-1\right)^2\frac{\rho_{max}}{\rho\chi_\theta},\label{Ray_NC_Config}
\end{equation}
where
\begin{equation}\label{zetatheta}
\zeta_\theta \equiv \frac{\rho_{max}}{\rho-\rho_{max}}\left[1-\frac{\rho_{max}^{nc}}{\rho_{max}}+2\frac{a\theta\lambda p_\phi(\rho_{max}^{nc}-\rho)^{1/2}}{\sqrt\rho}-\frac{\rho_{max}^{nc}-\rho_{max}}{\rho}\right],
\end{equation}
\begin{equation}
\xi_\theta\equiv \left[1+\frac{\rho-\rho_{max}^{nc}}{\rho_{max}}\right]\left[1+\left(\frac{\rho^{1/2}(\rho_{max}^{nc}-\rho)^{1/2}- \gamma\lambda(a\theta\lambda p_\phi)}{\rho+\rho_{max}-\rho_{max}^{nc}}\right)^2\right],
\end{equation}
\begin{equation}
\chi_\theta \equiv \left(\frac{\rho_{max}^{nc}-\rho_{m}}{\rho}-1\right)^2+ 
 \left(a\theta\lambda p_\phi\frac{\rho_{max}}{\rho}-\sqrt{\frac{\rho_{max}^{nc}}{\rho}-1}\right)^2.
\end{equation}
In the limit when $\theta\to0$ in Eqs. (\ref{Friedmann_NC_Config}) and (\ref{Ray_NC_Config}) have the correct limit, recovering the modified Friedmann equations (\ref{eff-FRW}) and (\ref{eff-RE}). In this setup, in contrast to section \ref{section_mom}, the viability of an inflationary period is null. In this case, there is no effective potential that could generate inflation.

\section{Discussion and Outlook}
In this work we have obtained the noncommutative version of the Friedmann equations for the flat FRW model with a free scalar field in the effective loop quantum cosmology scheme, following closely the construction made in Ref.~\refcite{primogenito}.\\
First, we introduce a deformation between the momentum sector variables $v$ and $p_{\phi}$. Analytical solutions to the noncommutative equations of motion (\ref{edm_momentos}) could be found. From the solution, for $v(t)$ it is shown that, even with the presence of noncommutativity, the bounce is preserved (which is one of the signature features of LQC), and after the bounce the volume has a slower growth, as depicted in Fig.~\ref{figura1a}. On the other hand, the solution for the scalar field shows that it is independent of the noncommutative parameter, as seen in Fig.~\ref{figura1b}, supporting the fact that the energy density profile is the same as in standard LQC (as previously pointed out employing different arguments in Ref.~\refcite{primogenito}). With the solutions of the equations of motion at hand, a noncommutative version of the Friedmann equation is constructed, given by Eq.~(\ref{Friedmann_NC_Mom}), and whose commutative limit ($\eta\to0$) is the correct one. Despite having all the pieces to construct the noncommutative version of Eq.~(\ref{eff-RE}), a compact form of such an equation could not be obtained. Nonetheless, it can be shown that when the limit $\eta\to0$ is taken, Eq.~(\ref{aceleracion_momentos}) also reduces to the commutative one, Eq.~(\ref{eff-RE}). Furthermore, we show that the presence of noncommutativity in the momentum sector induces an effective potential, within the free theory, opening the possibility of having an inflationary epoch. In Fig.~\ref{epsilon} we can see the accelerated evolution of the slow-roll parameter $\epsilon\equiv-\dot{\mathrm H}/\mathrm{H}^2$, which in an inflationary era must satisfy $\epsilon<1$; we can see that this condition is met, however, this is not a sufficient condition for inflation to occur. We, therefore, set to investigate the number of e-foldings produced immediately after the bounce. We found that inflation does not take place, since the number of e-foldings is barely close to one (and at least we need 60), as shown in Fig.~\ref{num_efoldings}. 
Finally, in the case of noncommutativity among the variables of the configuration sector, $\beta$ and $\phi$, it was also possible to find a noncommutative version of the modified Friedmann equation and the acceleration equation, given by the Eqs. (\ref{Friedmann_NC_Config}) and (\ref{Ray_NC_Config}), respectively. When the limit $\theta\to0$ is taken in Eq.~(\ref{Friedmann_NC_Config}) and Eq.~(\ref{Ray_NC_Config}), we get the correct commutative equations, given by Eq.~(\ref{eff-FRW}) and Eq.~(\ref{eff-RE}). In addition, unlike the previous case, the introduction of noncommutativity does not induce any effective potential, so an inflation period cannot be established.

\section*{Acknowledgments}
S.P.P. was partially supported by SNI-CONACyT, M\'exico. J.S. was partially funded by PRODEP grant UGTO-CA-3.


\end{document}